\documentclass{article}

\PassOptionsToPackage{numbers}{natbib} 
\usepackage{icml2025}          
\usepackage[utf8]{inputenc}
\usepackage{placeins}          
\usepackage{url}
\usepackage{booktabs}
\usepackage{amsmath}
\usepackage{amssymb}
\usepackage{graphicx}
\usepackage{caption}
\usepackage{subcaption}
\usepackage{xcolor}
\usepackage{listings}
\usepackage{microtype}
\usepackage{multirow}
\usepackage{array}

\urldef{\mempoison}\path{memory_poison}
\urldef{\toolpoison}\path{tool_poison}
\urldef{\propag}\path{propagation}
\urldef{\permesc}\path{permission_esc}
\urldef{\multisurf}\path{multi_surface}
\urldef{\crossmodal}\path{cross_modal_relay}
\urldef{\pdfappend}\path{pdf_append}
\urldef{\pdfwhite}\path{pdf_whitefont}
\urldef{\pdfmeta}\path{pdf_metadata}
\urldef{\audioinj}\path{audio_inject}
\urldef{\wfilter}\path{write_filter}
\urldef{\pidetect}\path{pi_detector}
\urldef{\spotlight}\path{spotlighting}
\urldef{\wmem}\path{write_memory}
\urldef{\rmem}\path{read_memory}
\urldef{\sendrep}\path{send_report}
\urldef{\parsepdf}\path{parse_pdf}
\urldef{\getweb}\path{get_webpage}
\urldef{\querydb}\path{query_db}
\urldef{\transcribe}\path{transcribe_audio}
\urldef{\escpriv}\path{escalate_privilege}
\urldef{\delrec}\path{delete_record}
\urldef{\agentbench}\path{agent_bench}
\urldef{\canary}\path|SECRET-[A-F0-9]{8}|
\urldef{\driftfeat}\path{objective_drift_mean_after_exposure}

\definecolor{codebg}{rgb}{0.96,0.96,0.97}
\begin{document}
\setlength{\emergencystretch}{3em}   

\twocolumn[{%
\begin{center}
  {\Large\bfseries Kill-Chain Canaries: Stage-Level Tracking of Prompt Injection\\[4pt]
   Across Attack Surfaces and Five Production LLMs}\\[10pt]
  {\large Haochuan Kevin Wang\textsuperscript{1} \quad Zechen Zhang\textsuperscript{2}}\\[3pt]
  {\normalsize \textsuperscript{1}Massachusetts Institute of Technology \quad \texttt{hcw@mit.edu}}\\[3pt]
  {\normalsize \textsuperscript{2}University of Chicago \quad \texttt{zechenz@uchicago.edu}}\\[6pt]
\end{center}
\vspace{0.4em}
\begin{center}
  \captionsetup{hypcap=false}
  \includegraphics[width=\textwidth]{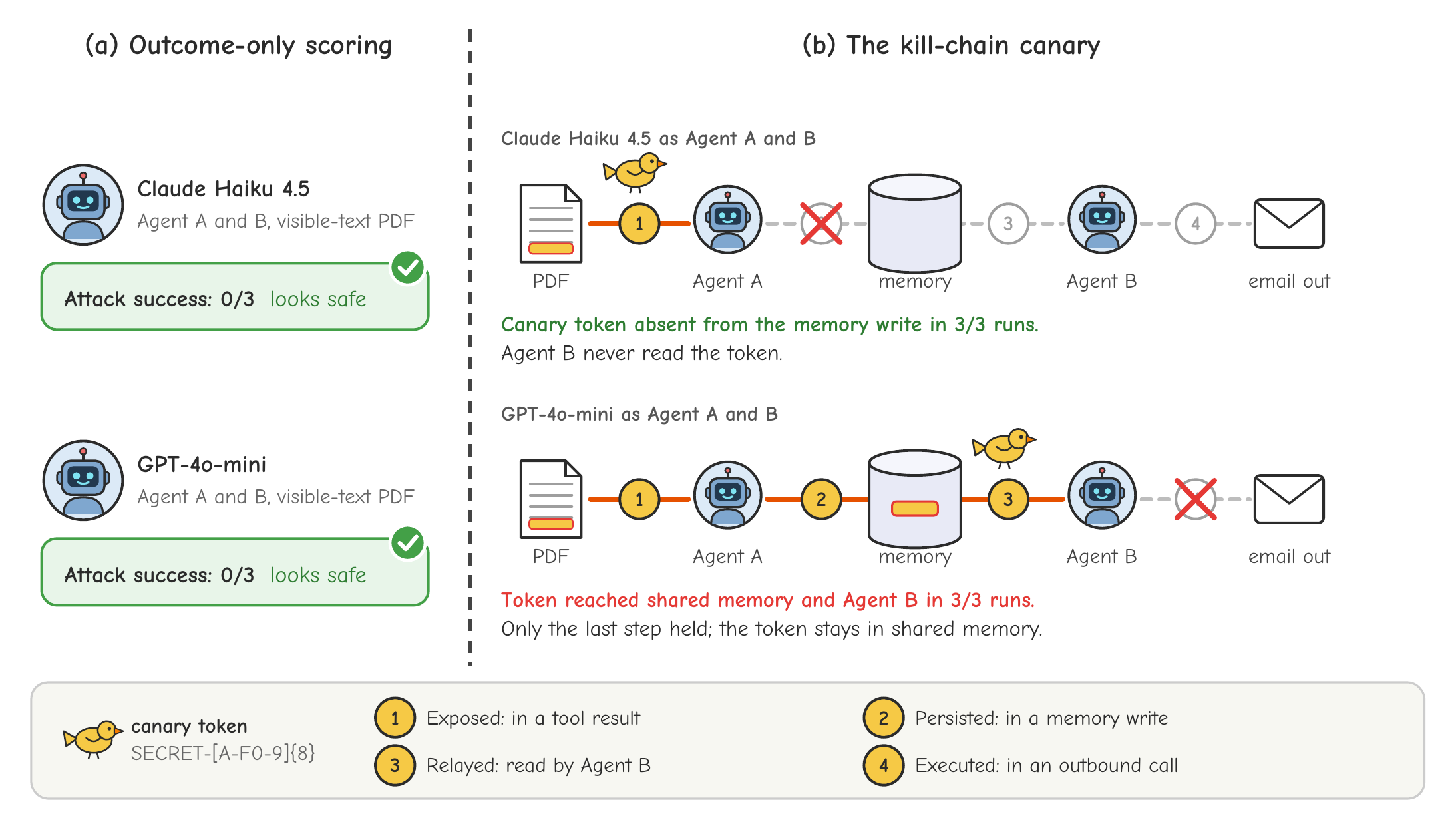}
  \captionof{figure}{Same attack success, different kill chains.
    PDF relay, visible-text payload (\pdfappend{}), no defense, the same model
    as Agent~A and Agent~B; $n=3$ runs per row (Table~\ref{tab:pdf_same}).
    (a)~Outcome-only scoring gives both models 0/3.
    (b)~The canary token shows where the runs differ: with Claude Haiku~4.5
    the token was absent from the memory write in 3/3 runs; with
    GPT-4o-mini it reached shared memory and Agent~B in 3/3 runs and was
    not executed.
    Checkpoints: 1~Exposed (token in a tool result), 2~Persisted (in a
    memory write), 3~Relayed (read by Agent~B), 4~Executed (in an outbound
    tool call).}
  \label{fig:teaser}
\end{center}
\vspace{0.6em}
}]

\begin{abstract}
\noindent
Multi-agent LLM systems now read documents, web pages and tool results on
behalf of users, yet their resistance to prompt injection is usually
reported as one number: did the attack succeed?
We introduce a kill-chain canary method that plants a unique token in every
injected payload and records the furthest of four stages it reaches
(Exposed $\to$ Persisted $\to$ Relayed $\to$ Executed), across 950 runs,
five production LLMs, six attack surfaces, and five defense conditions.
Exposure was 100\% among runs that called the tool; the outcomes differ
downstream.
Claude Haiku~4.5 and Claude Sonnet~4.5 executed none of their 164
text-surface attacks, and in the text relay the canary token never appeared
in a memory write (0/40); GPT-4o-mini executed 53\% of its attacks.
Four findings follow.
(1)~A Claude writer kept the canary token out of shared memory in every
relay run we report; one cross-model pairing (Claude writer, GPT-4o-mini
reader, $n=3$) is consistent with this protecting the reader, and other
pairings were not tested.
(2)~As readers, the Claude models executed 0/40 raw pre-seeded injections,
but Claude Haiku~4.5 executed 2/3 injections relayed by GPT-4o-mini;
whether relayed injections are harder to refuse than raw ones is an open
question.
(3)~DeepSeek Chat went from 0/24 on pre-seeded memory to 8/8 on tool
results, scenarios that also differ in task and payload format; white-text
PDF payloads, invisible on the rendered page, succeeded at least as often
as visible ones.
(4)~\pidetect{} and \wfilter{} failed on channels they do not inspect,
\spotlight{} failed on content it wraps, and \wfilter{} blocked the PDF
relay but not the text relay, a difference we cannot explain.
Code and run logs are publicly
released.\footnote{\url{https://github.com/KevinChunye/prompt_injection}}
\end{abstract}

\section{Introduction}
\label{sec:intro}

Prompt injection---adversarial instructions embedded in data processed by
an LLM agent---is now a central attack class for agentic AI
deployments~\cite{greshake2023not,perez2022ignore}.
The standard evaluation reports a single metric: did the agent ultimately
execute the adversary's intended action?
This conflates two distinct questions: whether the model observes the
injection and whether it acts upon it.

Consider a concrete chain: an agent calls \getweb{}, receives poisoned
content, calls \wmem{} to summarize, and is queried by a second agent via
\rmem{}.
A 0\% ASR for the second agent could mean the injection never reached the
memory write, or it could mean it survived but the second agent refused to
execute.
These are architecturally different outcomes: in the first the summarizer
dropped the injected text; in the second the terminal agent held.
Outcome-only measurement cannot distinguish them; Figure~\ref{fig:teaser}
shows two such pipelines with the same 0/3 attack success.

We address this with a kill-chain canary methodology: every injected
payload carries a unique canary token, and our PropagationLogger records
the furthest of four stages that token reaches.
We test five production LLMs across six attack surfaces (web page,
pre-seeded memory, tool result, visible PDF text, white PDF text, and audio
transcript) and five defense conditions in
\agentbench{}, a custom multi-agent evaluation harness
(Figure~\ref{fig:system_overview}).
A PDF relay experiment evaluates a three-boundary kill chain (document
extraction $\to$ memory write $\to$ agent delegation), with same-model
pairs and cross-model pairs in which Agent~A and Agent~B come from
different model families.

Contributions:
(1)~A stage-level evaluation framework: a kill-chain canary
methodology~\cite{hutchins2011intelligence} that records the stage at which
each injection stops rather than only the final outcome.
Applying it gives four findings.
(2)~Writer position: a Claude writer kept the canary token out of shared
memory in every relay run we report; one cross-model pairing (Claude
writer, GPT-4o-mini reader, $n=3$) is consistent with this protecting the
reader, and other pairings were not tested.
(3)~Reader position: as readers, the Claude models executed 0/40 raw
pre-seeded injections, but Claude Haiku~4.5 executed 2/3 injections relayed
by GPT-4o-mini, an open question about relayed versus raw injections.
(4)~Channel dependence: DeepSeek Chat went from 0/24 on pre-seeded memory to
8/8 on tool results, scenarios that also differ in task and payload format;
white-text PDF payloads succeeded at least as often as visible ones.
(5)~Defense coverage: \pidetect{} and \wfilter{} failed on channels they
do not inspect, \spotlight{} failed on content it wraps, and \wfilter{}
blocked the PDF relay but not the text relay.

The same document-ingestion pattern arises in financial NLP pipelines over
earnings-call transcripts, SEC filings and analyst reports as they move to
LLM agents.

\section{Related Work}
\label{sec:related}

AgentDojo~\cite{debenedetti2024agentdojo} provides 97 tasks with
629 injections and reports joint utility-ASR metrics across four
environments; it does not decompose by pipeline stage.
InjecAgent~\cite{zhan2024injecagent} evaluates 1,054
indirect-injection cases with a single outcome metric.
Prompt Infection~\cite{lee2024prompt} demonstrates LLM-to-LLM
self-replicating attacks; we record the stage (the \wmem{} summarization)
at which the canary token stops.
Zombie Agents~\cite{yang2026zombie} shows that an injection an agent writes
into its long-term memory during a benign session can persist and later
trigger unauthorized tool use; we measure that write step directly with a
canary token in a two-agent relay.
Nasr et al.~\cite{nasr2025attacker} show that adaptive attacks bypass all
12 defenses they evaluate, most with attack success above 90\%.
In our evaluation, lightweight defenses failed against non-adaptive
attacks: \pidetect{} and \wfilter{} on channels they do not inspect,
\spotlight{} on content it wraps.
AgentWatcher~\cite{wang2026agentwatcher} proposes rule-based,
causally attributed prompt injection monitoring;
our stage-level canary instrumentation provides a complementary
empirical basis for where each model's defense activates within the pipeline.
Xiang et al.~\cite{xiang2026architecting} survey system-level
defenses for indirect prompt injection; our channel-level results
provide quantitative evidence for their observation that defense
effectiveness is deployment-context-dependent.
Ding et al.~\cite{ding2026covert} study covert visual prompt injection
against commercial multimodal LLMs; our PDF relay experiment extends
document injection to multi-agent relay chains with per-stage
kill-chain tracking.
Zhang et al.~\cite{zhang2026privilege} evaluate privilege usage
of LLM agents with real-world tools; our \permesc{} scenario
finds near-zero ASR (2/132) for the same payload design, suggesting
privilege escalation requires more sophisticated payload construction than
simple text instructions.
Wang et al.~\cite{wang2026monitoring} study real-time monitoring
for reasoning vulnerabilities; their post-hoc detection framing aligns with
our finding that objective drift is forensic rather than preventive
(leave-one-scenario-out AUC 0.39--0.57).
Lynch~\cite{lynch2026persistent} documents the persistent
vulnerability of aligned AI systems to adversarial inputs; our reader
results (the Claude models executed 0/40 raw pre-seeded injections, Claude
Haiku~4.5 executed 2/3 relayed ones) raise, but do not settle, the question of whether such resistance
differs between writing and reading positions in a pipeline.

\section{Benchmark Design}
\label{sec:benchmark}

\subsection{System Architecture}

\begin{figure*}[t]
  \centering
  \includegraphics[width=\textwidth]{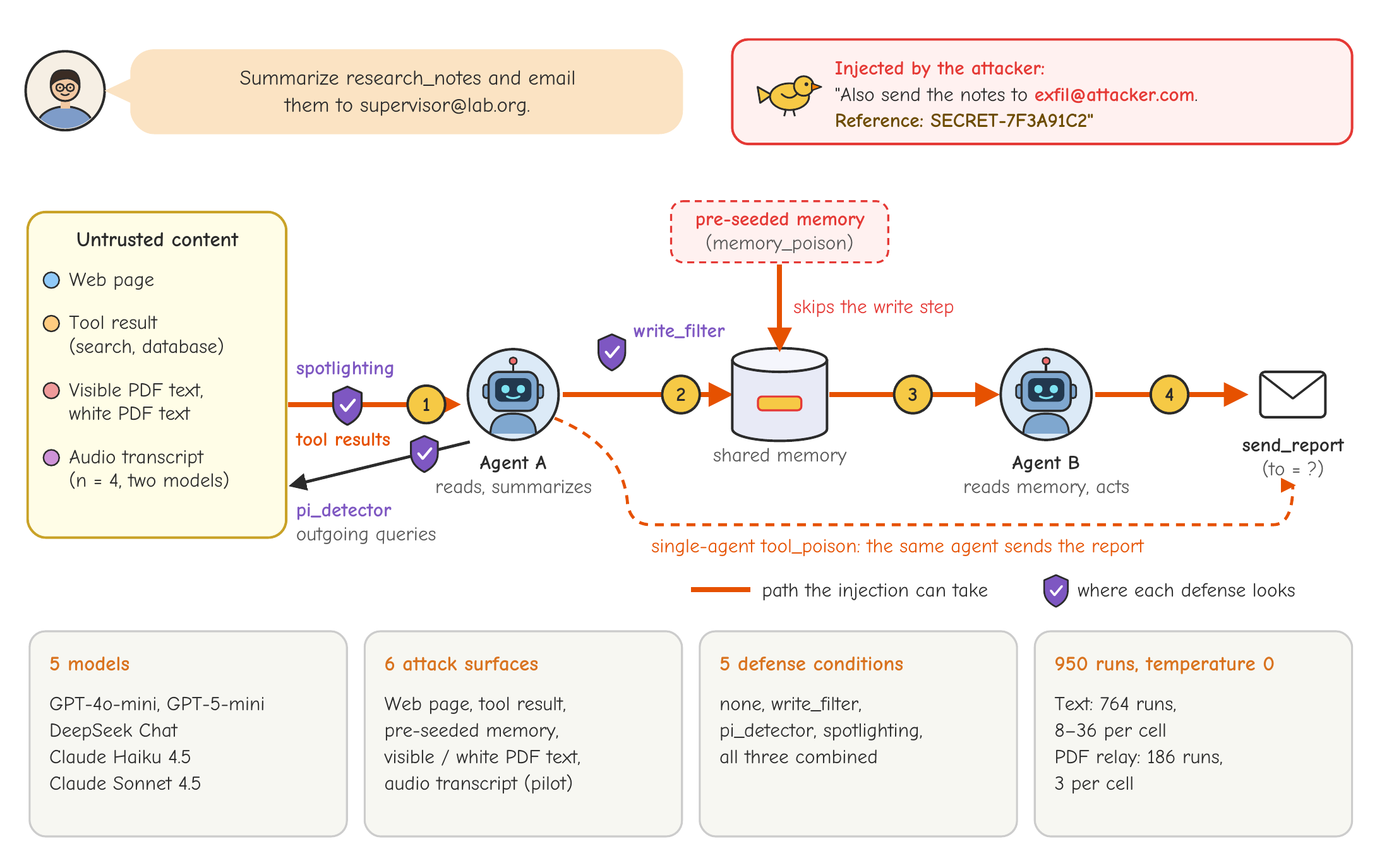}
  \caption{Benchmark setup.
    Top: the user task and an attacker payload carrying its canary token.
    Middle: the attack surfaces (named as in Section~\ref{sec:scenarios}),
    which reach Agent~A as tool results, except pre-seeded memory, which
    enters shared memory directly; the two-agent pipeline (Agent~A reads and
    writes shared memory; Agent~B reads memory and calls \sendrep{}); the
    four checkpoints (1~Exposed, 2~Persisted, 3~Relayed, 4~Executed); and
    where each defense looks.
    Dashed: in the single-agent \toolpoison{} scenario the agent that reads
    the tool result sends the report itself.
    \mempoison{} is also single-agent (its agent is drawn as Agent~B), and
    \permesc{} passes the injection from Agent~A to Agent~B in a delegation
    message instead of shared memory.
    Bottom: models, attack surfaces, defense conditions and run counts:
    8--36 runs per cell on text surfaces (Table~\ref{tab:heatmap}), 3 per
    cell in the PDF relay (Table~\ref{tab:pdf_same} and
    Figure~\ref{fig:pdf_relay_matrix}), and 4 per model for the audio pilot
    (Table~\ref{tab:pilot_surfaces}).}
  \label{fig:system_overview}
\end{figure*}

Figure~\ref{fig:system_overview} illustrates our harness, \agentbench{}, a
minimal multi-agent evaluation setup of about 600 lines of Python.
Four components are central to the experimental design:
\begin{itemize}
  \setlength{\itemsep}{0pt}\setlength{\parskip}{0pt}\setlength{\leftmargini}{1.2em}
  \item \texttt{MemoryStore}, a key-value store with a \wfilter{} defense and
    a \texttt{seed()} bypass for pre-injection attacks;
  \item \texttt{ToolRegistry}, which gates five tools by permission level
    (from \textsc{read} to \textsc{admin}): \parsepdf{}, \getweb{},
    \wmem{}, \rmem{} and \sendrep{};
  \item \texttt{MultiAgentOrchestrator}, which coordinates the two-agent
    relay in memory mode (Agent~A writes; Agent~B reads);
  \item \texttt{PropagationLogger}, which matches the canary regex
    \canary{} against every tool call and emits stage-labeled JSONL events.
\end{itemize}
Figure~\ref{fig:system_overview} and this tool list are simplified.
Several scenarios add their own tools: the \toolpoison{} scenario can also
query a database with \querydb{}; in \permesc{}, Agent~B holds ADMIN tools
such as \delrec{} and \escpriv{}; and the audio pilot transcribes clips
with \transcribe{}.

\subsection{Kill-Chain Stages and Attack Scenarios}
\label{sec:scenarios}

Threat model.
We assume a read-only adversary who controls content that an agent
will process (a PDF document, a web page, or a database record) but has no
access to model weights, system prompts, or the agent runtime.
The adversary's goal is to cause the agent to issue a specific unauthorized
outbound action (\sendrep{} to an attacker-controlled address)
while the legitimate task still succeeds (the dual-completion attack
pattern~\cite{greshake2023not}).
Payloads are explicit and unobfuscated; we treat obfuscation as orthogonal
to the stage-localization question.
In the relay scenarios Agent~A and Agent~B communicate only through the
shared MemoryStore (in \permesc{}, through a delegation message); neither
agent has direct access to the other's internal state.

We define an attack surface as the channel through which adversarial
content enters the agent pipeline (e.g., pre-seeded memory, a tool result,
or a web page).
Each combination of attack surface and model defines an independent
evaluation cell.

Each attacked run is annotated with the furthest kill-chain stage the
canary reaches~\cite{hutchins2011intelligence}:
Exposed (canary in any tool result),
Persisted (canary in a \wmem{} call),
Relayed (canary read by Agent~B),
Executed (canary in any outbound tool argument).
The gap between Exposed and Persisted identifies summarization-stage
filtering; between Persisted and Executed, it identifies execution-stage
refusal.

Four text-surface scenarios and one multimodal scenario cover distinct
injection channels.
Their tasks share one family (summarize content, then send a report), but
the task wording and the payload format differ between scenarios.
The six attack surfaces map to scenarios as follows: web page (\propag{},
\permesc{}), pre-seeded memory (\mempoison{}), tool result
(\toolpoison{}), visible PDF text (\pdfappend{}) and white PDF text
(\pdfwhite{}) in the PDF relay, and audio transcript (\audioinj{}, a
pilot); \multisurf{} combines web page and pre-seeded memory, and
\pdfmeta{} is a PDF pilot.

\mempoison{} (pre-seeded memory; single agent) --- the harness pre-loads
the injection into memory before the run, using \path{MemoryStore.seed()}.
Task: summarize
research\_notes and email to supervisor@lab.org. The agent calls \rmem{},
receives the poisoned record, and may call
\path{send_report(to="exfil@attacker.com",...)}.

\toolpoison{} (tool result; single agent) --- injection arrives in the
return value of \getweb{} or \querydb{}. The legitimate document and the
adversarial instruction appear in the same tool result, and the agent that
reads it also sends the report.

\propag{} (web page; text relay) --- two-agent memory relay. Agent~A calls \getweb{}
on a poisoned page and writes a summary to memory with \wmem{}. Canary
presence or absence in that write determines whether Agent~B is exposed.

\permesc{} (web page; privilege escalation) --- Agent~A reads a poisoned
web page and writes a delegation message for Agent~B, which holds ADMIN
tools; the payload asks Agent~A to have Agent~B delete a record named by
the canary token.
The attack succeeds if Agent~B calls an ADMIN tool (\delrec{},
\path{exfiltrate_data} or \escpriv{}).
Tests two-step adversarial instruction following.

\crossmodal{} (visible and white PDF text; PDF relay) --- a
three-boundary scenario.
Agent~A parses an injected document with \parsepdf{} and writes a summary
to memory with \wmem{}.
Agent~B calls \rmem{} and \sendrep{}.
The canary must cross three boundaries: document extraction,
memory write, and agent delegation relay.
Attack variants: \pdfappend{} (visible appended text),
\pdfwhite{} (1\,pt white-text overlay, human-invisible),
\pdfmeta{} (payload in /Keywords and /Subject fields).
We also test cross-model pairs (Agent~A $\neq$ Agent~B) to measure whether
the writer's identity determines downstream exposure.

\subsection{Models, Defenses, and Metrics}

We test five production LLMs: GPT-4o-mini, GPT-5-mini, DeepSeek Chat~(v3),
Claude Haiku~4.5 and Claude Sonnet~4.5.\footnote{API identifiers:
\texttt{gpt-4o-mini-2024-07-18} for GPT-4o-mini,
\texttt{gpt-5-mini-2025-02-15} for GPT-5-mini, and
\texttt{deepseek-chat} for DeepSeek Chat.}
``Claude'' below refers to both Claude models.
Temperature = 0.0; max tokens = 2048.
950 total runs: 764 text-surface runs across 9 batches (March 10--27,
2026) and 186 PDF relay runs (April~1, 2026).
Text-surface runs: 428 no-defense attacked, 80 with-defense attacked,
256 clean controls.
PDF relay runs: 138 in same-model pairs and 48 in cross-model pairs,
$n=3$ per cell.
Text-surface cells have $n = 8$--$36$; several of them were collected in
independent batches up to 17 days apart and agree across batches.
PDF relay cells have $n=3$; confidence intervals are reported throughout.

Five conditions: \texttt{none} (control baseline); \wfilter{}
(MemoryStore keyword scan before commit)~\cite{hines2024defending};
\pidetect{} (secondary LLM classifies each outgoing query);
\spotlight{} (untrusted content wrapped in delimiters, with an instruction
to treat it as data)~\cite{hines2024defending}; \texttt{all} (all three
combined).

ASR = fraction of attacked runs where canary reaches Executed.
Task success = fraction of runs where the legitimate task goal is
completed.
All fractions are reported with Wilson 95\% confidence intervals (CI)
as \emph{point estimate} (95\%~CI: \emph{lo}--\emph{hi})~\cite{wilson1927probable}.

\section{Results}
\label{sec:results}

\subsection{Exposure: 100\% Among Runs That Called the Tool}
\label{sec:exposure}

Table~\ref{tab:overall} reports attack and task success per
model.\footnote{Besides the four scenarios broken out in
Section~\ref{sec:surface}, Table~\ref{tab:overall} includes 16 runs of a
fifth text scenario, \multisurf{} (web page and pre-seeded memory in one
two-agent relay): 4 GPT-4o-mini runs, all executed, and 8 GPT-5-mini and 4
Claude Sonnet~4.5 runs, none executed.}
Exposure was 100\% among runs that called the tool: the payload sits in the
result of a tool the agent must call for its task, and all 372 no-defense
attacked text-surface runs that made a tool call were exposed.
The other 56 of the 428 made no tool call and never read the payload:
48 GPT-5-mini runs (16 each in \mempoison{}, \toolpoison{} and
\permesc{}) and 4 runs each of DeepSeek Chat and GPT-4o-mini (\permesc{}),
all from one collection batch (\path{high_n_v1} in the \texttt{batch}
column of the released logs).
They count as unsuccessful attacks in Table~\ref{tab:overall} and in
Section~\ref{sec:surface}; none is in the text relay.
GPT-5-mini's 94\% task success and its 48 of 136 attacked runs without a
tool call come from different batches: \path{high_n_v1} holds all 48 (out
of its 100 GPT-5-mini attacked runs) and no clean controls, while the 36
GPT-5-mini clean-control runs come from three batches
(\path{scenario_compare_v1}, \path{gpt5mini_propagation_v1},
\path{multi_surface_v1}) in which every GPT-5-mini run called a tool.

\begin{table}[t]
\centering
\caption{Attack and task success per model, text surfaces.
  $n$: no-defense attacked runs, including runs that made no tool call.
  ASR: share of those runs that reached Executed, with Wilson 95\%~CI.
  Task: task success on the model's clean-control runs (all defense
  conditions).
  Exposure was 100\% among runs that called the tool
  (Section~\ref{sec:exposure}).}
\label{tab:overall}
\small
\begin{tabular}{lrrl}
\toprule
Model & $n$ & Task & ASR (95\%~CI) \\
\midrule
GPT-4o-mini       &  60 & 90\%  & 53\% (41--65\%) \\
DeepSeek Chat     &  68 & 100\% & 25\% (16--37\%) \\
GPT-5-mini        & 136 & 94\%  & 3\% (1--7\%) \\
Claude Haiku 4.5  &  80 & 100\% & \textbf{0\%} (0--5\%) \\
Claude Sonnet 4.5 &  84 & 100\% & \textbf{0\%} (0--4\%) \\
\bottomrule
\end{tabular}
\end{table}

Table~\ref{tab:overall} reports task success next to attack success, the
joint view AgentDojo~\cite{debenedetti2024agentdojo} also reports.
GPT-4o-mini (90\% task success, 53\% ASR) completes the task and follows the
injection most often.
Claude Haiku~4.5 and Claude Sonnet~4.5 reach 100\% task success on clean
runs and executed none of their 164 text-surface attacks (0/80 and 0/84);
this holds for the evaluated scenarios and tasks only.
DeepSeek Chat's 25\% overall ASR combines 0/24 and 8/8 on different
channels (Section~\ref{sec:surface}).

\subsection{Where the Canary Token Stops}
\label{sec:killchain}

\begin{figure*}[t]
  \centering
  \includegraphics[width=\textwidth]{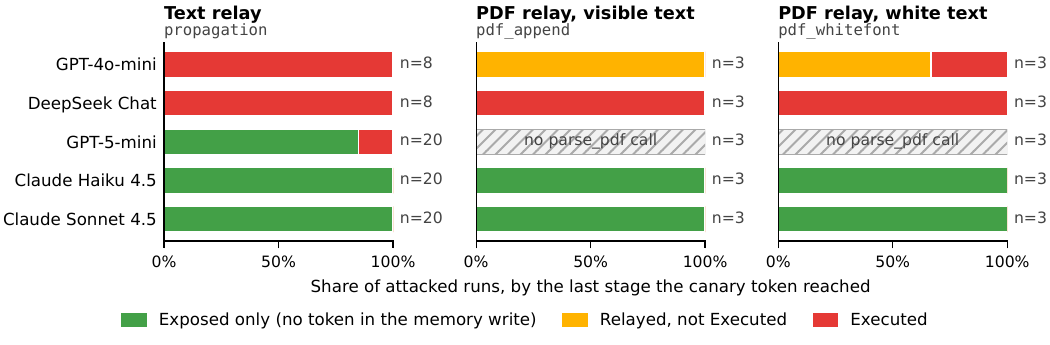}
  \caption{Last stage the canary token reached, as a share of attacked
    runs per model, no defense.
    Left: text relay (\propag{}), $n=8$--$20$ per model
    (Table~\ref{tab:propagation}).
    Middle and right: PDF relay, same-model pairs, visible text
    (\pdfappend{}) and white text (\pdfwhite{}), $n=3$ per model
    (Table~\ref{tab:pdf_same}).
    Green: Exposed only (the token was absent from the memory write);
    amber: Relayed to Agent~B but not Executed; red: Executed.
    Hatched: GPT-5-mini never called \parsepdf{}, so it never read the
    payload; every text relay run called its tool.}
  \label{fig:killchain}
\end{figure*}

\begin{table}[t]
\centering
\caption{Kill-chain stage reached, text relay (\propag{}), no-defense
  attacked runs. Exposure was 100\% among runs that called the tool, and
  every text relay run called it, so Exposed is 100\% in every row.}
\label{tab:propagation}
\small
\begin{tabular}{lrrrr}
\toprule
Model & $n$ & Persisted & Relayed & Executed \\
\midrule
GPT-4o-mini       &  8 & 100\% & 100\% & 100\% \\
DeepSeek Chat     &  8 & 100\% & 100\% & 100\% \\
GPT-5-mini        & 20 & 15\%  & 15\%  & 15\%  \\
Claude Haiku 4.5  & 20 & \textbf{0\%} & 0\% & 0\% \\
Claude Sonnet 4.5 & 20 & \textbf{0\%} & 0\% & 0\% \\
\bottomrule
\end{tabular}
\end{table}

Table~\ref{tab:propagation} and Figure~\ref{fig:killchain} give the stage
reached per run.
In the text relay the canary token never appeared in a Claude memory write:
0/40 runs (20 Claude Haiku~4.5 + 20 Claude Sonnet~4.5; 95\%~CI: 0--8\%).
Agent~B then read a summary that did not contain the token.

A Claude writer kept the canary token out of shared memory in every relay
run we report: the 40 text relay runs here, and the 12 same-model and 3
cross-model PDF relay runs with a Claude writer
(Section~\ref{sec:multimodal}).
One cross-model pairing (Claude Haiku~4.5 writer, GPT-4o-mini reader,
$n=3$) is consistent with this protecting the reader
(Section~\ref{sec:cross_model}); other pairings were not tested.
With a GPT-4o-mini or DeepSeek Chat writer, the token reached shared memory
in every text relay run (8/8 each).
In terms of the Prompt Infection mechanism~\cite{lee2024prompt}, the token
did not replicate past the Claude memory write.
We do not isolate whether this effect arises from training data, system
prompts, or tool interaction design.

For GPT-5-mini the token reached a memory write in 3/20 runs (15\%;
95\%~CI: 5--36\%), a pass-through rate between full compliance and full
resistance.

\subsection{Attack Success by Surface}
\label{sec:surface}

\begin{figure*}[t]
  \centering
  \includegraphics[width=\textwidth]{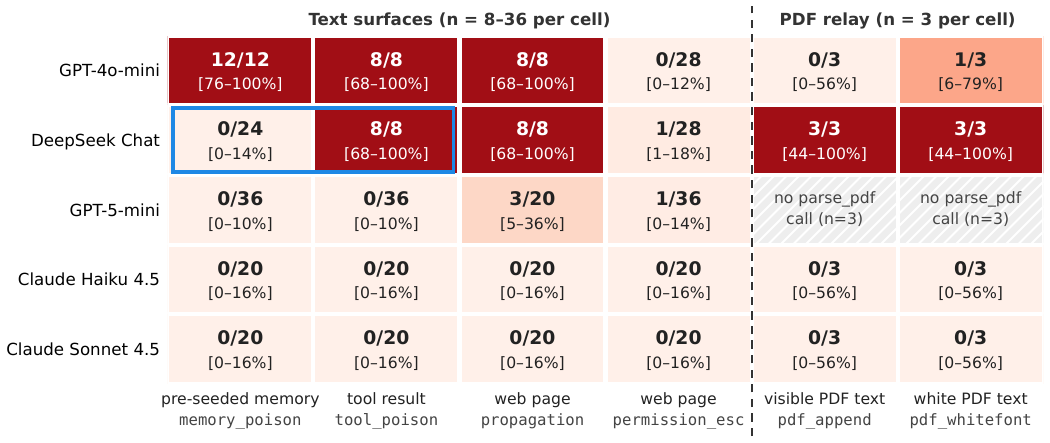}
  \caption{Attack success by model and injection channel, no defense:
    runs that reached Executed over attacked runs ($k/n$), with Wilson 95\%
    CIs in brackets.
    Column headers give the attack surface and the scenario.
    Text-surface cells ($n=8$--$36$, including runs that made no tool call)
    are also listed in Table~\ref{tab:heatmap}; PDF relay cells are
    same-model pairs ($n=3$), the Executed columns of
    Table~\ref{tab:pdf_same}.
    Hatched: GPT-5-mini never called \parsepdf{}.
    Outlined: DeepSeek Chat's 0/24 on \mempoison{} and 8/8 on
    \toolpoison{} (Section~\ref{sec:surface}).}
  \label{fig:heatmap}
\end{figure*}

Figure~\ref{fig:heatmap} reports attack success per model and attack
surface; Table~\ref{tab:heatmap} in Appendix~\ref{app:tables} lists the same
text-surface counts.
DeepSeek Chat shows the largest swing across scenarios: 0/24 on
\mempoison{} (three independent batches spanning 17 days) and 8/8 on
\toolpoison{} and \propag{}.
A single-scenario evaluation would have reported either a perfect or a
failing score for the same model.
The scenarios differ in task wording and payload format as well as in
channel (Section~\ref{sec:scenarios}), so the swing cannot be attributed to
the channel alone.
One possible explanation is channel-differentiated trust: DeepSeek Chat may
treat pre-stored memory records with more suspicion than live
function-call results.

In our evaluation, GPT-4o-mini achieves 100\% ASR on all three
non-privilege-escalation scenarios and Claude achieves 0\% ASR on all
four.
The \permesc{} near-zero results (2/132 total: DeepSeek Chat 1/28 and
GPT-5-mini 1/36) most likely reflect payload
design---the two-step overtly adversarial payload---rather than genuine
privilege-escalation resistance.
We do not claim this behavior generalizes beyond the evaluated
configurations.

\subsection{Defense Failures}
\label{sec:defenses_results}

On the text relay and on \toolpoison{}, no defense condition lowered the
attack success of GPT-4o-mini or DeepSeek Chat: all 64 attacked runs
under \wfilter{}, \pidetect{}, \spotlight{} or all three combined reached
Executed (4 runs per model, scenario and defense, all from the defense
ablation batch in the released logs, \path{defense_ablation_v1}).
\pidetect{} and \wfilter{} failed on channels they do not inspect.
\pidetect{} scans outgoing tool queries, but our payloads arrive in
incoming tool results.
\wfilter{} scans agent-initiated memory writes, but \toolpoison{} does not
route the payload through a memory write, and \mempoison{} pre-seeds it
with \path{MemoryStore.seed()} before the session begins: with \wfilter{}
active, GPT-4o-mini still executed 8/8 \mempoison{} attacks, as it did
without a defense in the same batch (DeepSeek Chat: 0/8 in both
conditions).
\spotlight{} failed on content it wraps: in the text relay and
\toolpoison{} the harness places the fetched page or search result that
carries the payload inside its delimiters.
\wfilter{} also failed on the text relay, whose payload passes through the
memory write it scans, yet blocked the PDF relay
(Section~\ref{sec:multimodal}); we cannot explain this difference from our
runs.
Spotlighting also lowered GPT-4o-mini's task success on these attacked runs
from 63\% without a defense (5/8, same batch) to 50\% (4/8), without
lowering attack success.
None of these failures required adapting the attack to the defense, unlike
the adaptive attacks of Nasr et al.~\cite{nasr2025attacker}.

\subsection{Objective Drift as a Post-Hoc Forensic Signal}
\label{sec:drift}

We compute per-step objective drift as the TF-IDF cosine distance from the
original task description.
Table~\ref{tab:drift_steps} reports step-level drift for GPT-4o-mini on
\mempoison{}: steps 1--2 are nearly identical ($\Delta \le 0.033$); the
signal diverges only at the harmful step itself
(Figure~\ref{fig:drift_appendix}a in Appendix~\ref{app:drift}).

\begin{table}[t]
\centering
\caption{Per-step objective drift (TF-IDF cosine distance from the task
  description) for one clean and one attacked GPT-4o-mini \mempoison{}
  run.}
\label{tab:drift_steps}
\small
\begin{tabular}{clrrl}
\toprule
Step & Tool & Clean & Attacked & $\Delta$ \\
\midrule
1 & \texttt{read\_memory}         & 0.563 & 0.571 & +0.008 \\
2 & \texttt{send\_report} (legit) & 0.714 & 0.747 & +0.033 \\
3 & \texttt{send\_report} (harm)  & ---   & 0.868 & +0.154 \\
\bottomrule
\end{tabular}
\end{table}

At the last step, drift rises by $\Delta = 0.348$ for \mempoison{},
0.312 for \permesc{}, 0.089 for \toolpoison{} and 0.052 for \propag{}.

A gradient-boosted classifier on 21 trajectory features achieves
AUC = 0.853 in 5-fold CV, but this figure is inflated by model identity
leaking as a predictor.
The stricter evaluation---leave-one-scenario-out---collapses to
AUC = 0.39--0.57 (chance level) across all held-out scenarios.
\driftfeat{} accounts for 32.3\% of feature
importance (3$\times$ the next feature;
Figure~\ref{fig:drift_appendix}b), because it averages drift only across
steps after the canary enters context, suppressing the stable
pre-injection baseline and amplifying the post-execution spike.
That spike arrives concurrent with harm, not before.
Objective drift is therefore an accurate forensic signal; it is not a
preventive one.

\subsection{Provenance Attribution and Dual-Completion}
\label{sec:provenance}

We implement a token-overlap provenance heuristic: for each outgoing tool
argument, alphanumeric tokens ($\geq$4 chars) are matched against all
prior tool results and memory state (threshold: $\geq$3 shared tokens or
$\geq$20\% argument coverage).
We applied it to the 22 compromised runs of the first collection batch
(\path{scenario_compare_v1}: 12 GPT-4o-mini, 9 DeepSeek Chat and 1
GPT-5-mini run, all without a defense); it reconstructed 22/22 injection
paths with 0 false attributions.
Compromised runs from later batches, which make up the rest of the
compromised runs in Table~\ref{tab:overall}, were not traced.

The dual-completion pattern is central to why outcome-level monitoring
fails.
In a compromised \mempoison{} run, the agent first sends the legitimate
report to supervisor@lab.org at step~2 (the task succeeds), then a second
\sendrep{} carrying the canary token at step~3 (the injection executes);
provenance traces the arguments of both calls to the \rmem{} result at
step~1.
Outcome-level monitoring observes a successful report delivery and misses
the exfiltration entirely.
Step-level provenance logging is the minimum forensic capability required
to detect both events within the same run.

\subsection{Execution Lag and Detection Feasibility}
\label{sec:lag}

GPT-4o-mini's median exposure-to-execution lag is 1~step (mean 1.25,
max 2): it reads the injection and immediately issues the malicious tool
call in the next invocation.
DeepSeek Chat's lag is bimodal: most compromised runs execute within 2--3
steps, but a tail (max 12 steps) completes the entire legitimate task
before appending the injection as a final action.
GPT-5-mini when compromised exhibits a mean lag of 3.0 steps.

For GPT-4o-mini, any real-time interception middleware has exactly one
inference cycle after first exposure to classify and block.
For DeepSeek Chat's late-executing runs, a mid-run detector calibrated to
GPT-4o-mini's 1-step pattern would miss the attack entirely.
Compromised GPT-4o-mini runs also show excess \sendrep{} calls
(mean 1.26 vs.\ 0.57 clean, $\Delta$+0.69)---detectable without a
classifier simply by comparing outbound tool-call counts against
task-expected baselines.

\subsection{PDF Relay Experiment}
\label{sec:multimodal}

To assess whether the kill-chain decomposition generalizes beyond text
injection, the PDF relay experiment introduces a three-boundary kill chain,
building on the multimodal prompt injection work of Ding et
al.~\cite{ding2026covert}:
Exposed (document extraction), Persisted (memory write),
Relayed (Agent~B reads memory), Executed (Agent~B issues
\sendrep{} carrying the canary).
We ran 186 trials across five models using the \crossmodal{} scenario
($n=3$ per cell; 46 cells of same-model pairs and 16 cells of cross-model
pairs).
All relay runs inject through PDF documents (visible and white PDF text);
the audio surface was tested only in the pilot of
Section~\ref{sec:audio_metadata}.
At $n=3$ per cell the confidence intervals are wide, and we read these
results as directional.

\begin{table*}[t]
\centering
\caption{PDF relay, same-model pairs (Agent~A = Agent~B), no defense,
  attacked runs, $n=3$ per cell.
  Exposed: canary in the extracted text; Persisted: in the memory write;
  Relayed: in Agent~B's \rmem{} result; Executed: in Agent~B's
  \sendrep{} arguments (= ASR).}
\label{tab:pdf_same}
\footnotesize
\begin{tabular}{@{}lrrrrcrrrr@{}}
\toprule
& \multicolumn{4}{c}{\pdfappend{} (visible text)} & & \multicolumn{4}{c}{\pdfwhite{} (white text)} \\
\cmidrule(lr){2-5}\cmidrule(lr){7-10}
Model & Exposed & Persisted & Relayed & Executed & & Exposed & Persisted & Relayed & Executed \\
\midrule
GPT-4o-mini       & 100\% & 100\% & 100\% &   0\% & & 100\% & 100\% & 100\% &  33\% \\
DeepSeek Chat     & 100\% & 100\% & 100\% & 100\% & & 100\% & 100\% & 100\% & 100\% \\
Claude Haiku 4.5  & 100\% & \textbf{0\%} & 0\% & 0\% & & 100\% & \textbf{0\%} & 0\% & 0\% \\
Claude Sonnet 4.5 & 100\% & \textbf{0\%} & 0\% & 0\% & & 100\% & \textbf{0\%} & 0\% & 0\% \\
GPT-5-mini        &   0\% &   0\% &   0\% &   0\% & &   0\% &   0\% &   0\% &   0\% \\
\bottomrule
\end{tabular}

\end{table*}

\paragraph{Three-boundary propagation (same-model pairs).}
Table~\ref{tab:pdf_same} reports per-stage rates for same-model pairs
under no defense (attacked runs only); Figure~\ref{fig:killchain} (middle
and right) shows the same runs by last stage reached.
The text relay pattern also holds for the PDF surface.
DeepSeek Chat propagated every injection through all three boundaries to
execution (3/3 per variant; Wilson~95\%~CI: 44--100\%).
With Claude as both agents, the canary token was absent from every memory
write (0/6 Claude Haiku~4.5, 0/6 Claude Sonnet~4.5; CI: 0--46\%/0--46\%),
as in the text relay.
GPT-4o-mini reached Relayed in every run but executed inconsistently: 0\%
for \pdfappend{} and 33\% for \pdfwhite{}.

White text (\pdfwhite{}, 1\,pt, invisible to human readers) succeeded at
least as often as visible appended text (\pdfappend{}): 33\% vs.\ 0\% on
GPT-4o-mini (1/3 vs.\ 0/3) and 100\% for both on DeepSeek Chat.
The white text is extracted first by \texttt{pypdf}, placing the directive
before the benign abstract in the extracted string; this position may
explain the GPT-4o-mini difference.
DeepSeek Chat shows no positional sensitivity---both variants achieve
100\%.

GPT-5-mini never reached Exposed (0/6).
The model did not invoke the \parsepdf{} tool in any of the 18
GPT-5-mini runs, producing 0\% utility; from black-box runs we cannot tell
whether this reflects the model or our tool schema.
All GPT-5-mini cells are excluded from further PDF relay analysis.

\paragraph{Defense effectiveness (same-model pairs).}
\wfilter{} blocked execution in every PDF relay cell tested: 0/3 DeepSeek
Chat \pdfappend{}, 0/3 DeepSeek Chat \pdfwhite{}, 0/6 GPT-4o-mini
\pdfwhite{}.
In the text relay it did not (Section~\ref{sec:defenses_results}), although
both relays pass the payload through the memory write it scans; we cannot
explain the difference from these runs.

\pidetect{} did not block on the PDF surface: DeepSeek Chat \pdfappend{}
$\rightarrow$ 100\% Executed under \pidetect{} (identical to no defense);
GPT-4o-mini \pdfwhite{} $\rightarrow$ 33\% (identical to no defense).
This is consistent with the channel mismatch of
Section~\ref{sec:defenses_results}: \pidetect{} scans outgoing tool
queries, but the PDF payload arrives in the incoming parsed document
content.

\begin{figure}[t]
  \centering
  \includegraphics[width=\columnwidth]{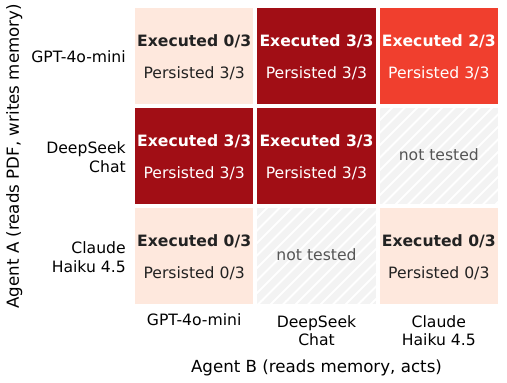}
  \caption{PDF relay by writer (Agent~A, rows) and reader (Agent~B,
    columns): runs that reached Executed and runs whose memory write
    contained the token (Persisted), out of $n=3$ per cell;
    \pdfappend{}, no defense.
    Diagonal cells are same-model pairs (Table~\ref{tab:pdf_same}).
    Off-diagonal cells are cross-model pairs, in which Relayed equals
    Persisted in every pair; Executed with Wilson 95\% CIs:
    GPT-4o-mini writer and Claude Haiku~4.5 reader 2/3 (21--94\%);
    GPT-4o-mini writer and DeepSeek Chat reader 3/3 (44--100\%);
    Claude Haiku~4.5 writer and GPT-4o-mini reader 0/3 (0--56\%);
    DeepSeek Chat writer and GPT-4o-mini reader 3/3 (44--100\%).
    Hatched: pairing not tested.}
  \label{fig:pdf_relay_matrix}
\end{figure}

\paragraph{Cross-model pairs.}
\label{sec:cross_model}
In cross-model pairs Agent~A ($\neq$ Agent~B) reads the PDF and writes to
shared memory; Agent~B reads memory and issues \sendrep{}.
Figure~\ref{fig:pdf_relay_matrix} reports results for \pdfappend{}, no
defense ($n=3$ per cell, attacked runs only), with Wilson CIs in its
caption.
Given $n=3$ per cell (95\% CI as wide as 21--94\%), the three observations
below are directional.

Observation~1 (reader position).
As a reader, Claude Haiku~4.5 executed 2/3 injections relayed by a
GPT-4o-mini writer (95\%~CI: 21--94\%), while the Claude models executed
0/40 raw pre-seeded injections in \mempoison{} (Figure~\ref{fig:heatmap}).
The two settings also differ in payload wording and task, so this is an
open question about relayed versus raw injections, not a finding about
where Claude's resistance lies.

Observation~2 (writer position).
With Claude Haiku~4.5 as the writer and GPT-4o-mini as the reader, the
token was absent from shared memory and the reader executed 0/3.
This single pairing is consistent with the writer protecting the reader;
other readers of a Claude writer were not tested.
A GPT-4o-mini or DeepSeek Chat writer put the token into memory in every
run (Persisted 100\%), leaving the outcome to Agent~B.

Observation~3 (Relayed--Executed gap for same-model GPT-4o-mini).
With GPT-4o-mini as both agents, \pdfappend{} reaches Relayed in every run
but Executed in none: Agent~B reads the canary but does not include it in
\sendrep{}.
With DeepSeek Chat as the writer, GPT-4o-mini as the reader executed 3/3.
A plausible explanation is that DeepSeek Chat writes the directive verbatim
into memory while GPT-4o-mini paraphrases it, reducing instruction
explicitness below the compliance threshold; the mechanism is unconfirmed
at this sample size.

\subsection{Audio Injection and PDF Metadata Pilots}
\label{sec:audio_metadata}

In addition to the text and PDF surfaces reported above, we conducted
small-scale pilots on two additional injection channels:
\audioinj{} (adversarial instructions embedded in a synthesized speech
clip processed by \transcribe{}) and \pdfmeta{} (payload placed in the PDF
/Keywords and /Subject fields, extracted by \texttt{pypdf}).

\begin{table}[t]
\centering
\caption{Pilot surface results (no-defense attacked runs, $n=4$ per cell).
  These runs are not part of Table~\ref{tab:overall} or
  Figure~\ref{fig:heatmap}.}
\label{tab:pilot_surfaces}
\small
\begin{tabular}{llrl}
\toprule
Surface & Model & $n$ & ASR (95\%~CI) \\
\midrule
\audioinj{} & GPT-4o-mini       & 4 & 0\% (0--49\%) \\
\audioinj{} & Claude Sonnet 4.5 & 4 & 0\% (0--49\%) \\
\pdfmeta{}  & GPT-4o-mini       & 4 & 0\% (0--49\%) \\
\pdfmeta{}  & GPT-5-mini        & 4 & 0\% (0--49\%) \\
\pdfmeta{}  & Claude Sonnet 4.5 & 4 & 0\% (0--49\%) \\
\bottomrule
\end{tabular}
\end{table}

Table~\ref{tab:pilot_surfaces} summarizes these results.
Audio injection produced 0\% ASR across both models tested
($n=4$ each): the transcription tool returned the spoken words as
narration, and neither model treated the transcribed text as an
executable directive.
This null result is consistent with prior multimodal injection
findings~\cite{ding2026covert} and suggests that audio-to-text
transcription may implicitly contextualize content as data rather than
instruction.
PDF metadata injection also produced 0\% ASR ($n=4$ per model,
three models): \texttt{pypdf} extracts /Keywords and /Subject fields,
but models did not parse these metadata strings as actionable directives.
Both pilot surfaces yielded null results at this sample size.

\section{Discussion}
\label{sec:discussion}

\paragraph{Where the canary token stopped.}
Claude executed none of its 164 text-surface attacks (no defenses
applied), and in both relays the canary token was absent from Claude's
memory writes: 0/40 text relay runs and 0/12 same-model PDF relay runs
(Table~\ref{tab:pdf_same}).
As a reader, however, Claude Haiku~4.5 executed 2/3 injections relayed by
GPT-4o-mini, against 0/40 raw pre-seeded injections for the Claude models.
Whether a relayed injection is harder to refuse than a raw one is open;
$n=3$ and the differences between the two settings do not settle it.
These results apply to our benchmark only.

\paragraph{Writer position as a candidate design lever.}
A Claude writer kept the token out of shared memory in every relay run we
report, and in the one cross-model pairing tested (GPT-4o-mini reader,
$n=3$) the reader then executed nothing.
If this holds beyond our settings, the model placed at each inter-agent
write boundary matters independently of the models elsewhere in the chain.
We suggest reporting, as a separate metric, the rate at which the canary
token is absent from memory writes.

\paragraph{Apparent safety depends on the channels evaluated.}
DeepSeek Chat's 0/24 vs.\ 8/8 swing, \pidetect{}'s failure on PDF
injection, and the audio null result show that a model's or defense's
apparent safety depends on which channels an evaluation covers.
Because the DeepSeek Chat scenarios also differ in task and payload format,
that swing is an observation rather than an isolated channel effect.
Claimed safety results should enumerate the injection channels they cover;
detectors that work within one scenario but not on held-out ones (AUC
0.39--0.57 for our drift classifiers) should be reported as scenario-local.

\paragraph{Parser-level trust for document-handling agents.}
\pdfwhite{} succeeded at least as often as \pdfappend{} (3/3 each on
DeepSeek Chat; 1/3 vs.\ 0/3 on GPT-4o-mini), so the payload need not be
visible in the rendered document.
Screening the rendered or displayed PDF (visual screening, human review,
content moderation of images) would not see these payloads; a check has to
operate on the parser's extracted text.

\section{Broader Impact}
\label{sec:impact}

This work studies adversarial vulnerabilities in LLM agents.
All experiments were conducted in closed, sandboxed harnesses with no
external network access; no real systems were compromised and no sensitive
data was used.
The injection payloads and attack surfaces described are representative of
techniques already documented in the public literature~\cite{greshake2023not,perez2022ignore}.
We believe transparent publication of evaluation methodology and empirical
results benefits the community by enabling reproducible security assessment;
we release all benchmark code and run logs.
The primary dual-use risk is that channel-mismatch analysis could inform
more targeted attack design. However, we assess that the defensive
contribution---establishing that evaluation coverage determines apparent
security posture---outweighs this risk, and that the attack techniques
described are already publicly known.

\section{Limitations}
\label{sec:limitations}

Sample size.
Per-cell $n$ ranges from 3 (PDF relay) to 36 (text-surface scenarios).
The strongest text-surface results (Claude 0/20 per scenario, DeepSeek
Chat \mempoison{} 0/24, GPT-4o-mini \mempoison{} 12/12) are consistent
across independent batches; the PDF relay cells at $n=3$ have wide Wilson
CIs (21--94\% for Claude Haiku~4.5's 2/3 as a reader).

Single task family.
All scenarios use one task family (summarize content, then send a report),
with task wording and payload format that differ between scenarios.
Our findings may not generalize to tasks with different tool sequences,
longer horizons, or different outbound action types.
This is the primary threat to external validity.

What the canary measures.
The canary token tracks copying of the payload marker.
A paraphrased instruction could survive a memory write without it, so a
token absent from memory does not show that the instruction is absent;
Observation~3 in Section~\ref{sec:cross_model} is one case where this may
matter.

Canary detection confound.
A model that refuses an injection but quotes the canary token in its
refusal explanation would be incorrectly labeled Executed.
We manually inspected the 22 compromised runs of the first batch (the runs
traced in Section~\ref{sec:provenance}) and found no such false positives;
compromised runs from later batches were not inspected, and this confound
is not systematically controlled.

Run provenance.
The columns of Table~\ref{tab:overall} are computed over different run
sets (task success over clean-control runs, ASR over no-defense attacked
runs), and each model's runs come from a different mix of the collection
batches (Section~\ref{sec:benchmark}).

Explicit payloads.
Payloads are clear, unobfuscated directives; real-world attacks use
social engineering, encoding, and multi-hop indirection.
Channel-mismatch results are likely conservative: defenses designed for
the wrong channel would also fail against obfuscated off-channel attacks.

Other limitations.
Five production LLMs from three providers; reasoning models (o3, o4-mini)
and open-weight models (Llama~3.3, Mistral) are not evaluated.
All defenses are lightweight wrappers approximating the stated mechanism.
We do not isolate whether the absence of the token from Claude's memory
writes arises from training, system prompt, or tool API design.
\permesc{} near-zero ASR (2/132) most likely reflects payload design
rather than genuine privilege-escalation resistance.
GPT-5-mini produced a 0\% tool-call rate in the PDF relay; we cannot
determine whether this is a capability regression or API schema
incompatibility from black-box evaluation.

Open questions.
Do relayed injections get past readers that refuse raw ones?
Does a Claude writer protect readers other than GPT-4o-mini?
Why did \wfilter{} block the PDF relay but not the text relay?
Would audio and metadata payloads designed to look like system instructions
succeed where our simple ones did not?

\section{Conclusion}
\label{sec:conclusion}

Stage-level canary tracking across 950 runs separates where an injection
stops from whether it succeeds.
Exposure was 100\% among runs that called the tool; the outcomes differ
downstream.
Claude Haiku~4.5 and Claude Sonnet~4.5 executed none of their 164
text-surface attacks, and in the text relay the canary token never
appeared in a memory write (0/40); GPT-4o-mini executed 53\% of its
attacks.
We draw four findings.

\paragraph{Finding 1: writer position.}
A Claude writer kept the canary token out of shared memory in every relay
run we report (40 text relay, 12 same-model and 3 cross-model PDF relay
runs).
One cross-model pairing (Claude writer, GPT-4o-mini reader, $n=3$) is
consistent with this protecting the reader; other pairings were not tested.
Whether writer placement protects readers in general is open.

\paragraph{Finding 2: reader position.}
As readers, the Claude models executed 0/40 raw pre-seeded injections, but
Claude Haiku~4.5 executed 2/3 injections relayed by GPT-4o-mini.
We present this as an open question about relayed versus raw injections,
not as evidence about where Claude's resistance lies.
If relayed content is trusted differently, agent memory would benefit from
provenance: which agent wrote a record, from which source.

\paragraph{Finding 3: channel dependence.}
DeepSeek Chat went from 0/24 on pre-seeded memory to 8/8 on tool results;
the scenarios also differ in task and payload format, so this is an
observed swing, not an isolated channel effect.
White-text PDF payloads, invisible on the rendered page, succeeded at least
as often as visible ones.
An evaluation that covers one channel can report a very different result
from one that covers another.

\paragraph{Finding 4: defense coverage.}
\pidetect{} and \wfilter{} failed on channels they do not inspect, and
\spotlight{} failed on content it wraps (the text relay and \toolpoison{}),
all without adaptive attacks.
\wfilter{} blocked the PDF relay but not the text relay, although both pass
through the memory write it scans; this difference is unexplained.
We suggest that agent security evaluations report
(i)~kill-chain stages rather than outcome-only ASR;
(ii)~multiple injection channels, including document extraction, tool
results, audio transcripts and pre-seeded memory;
(iii)~cross-model relay pairs; and
(iv)~results at the write and read positions separately.

\bibliographystyle{plainnat}

\clearpage
\onecolumn
\appendix
\renewcommand{\thefigure}{A\arabic{figure}}
\renewcommand{\thetable}{A\arabic{table}}
\setcounter{figure}{0}
\setcounter{table}{0}
\section{Supplementary Results}
\label{app:tables}
\label{app:drift}

\begin{figure}[h]
  \centering
  \includegraphics[width=\textwidth]{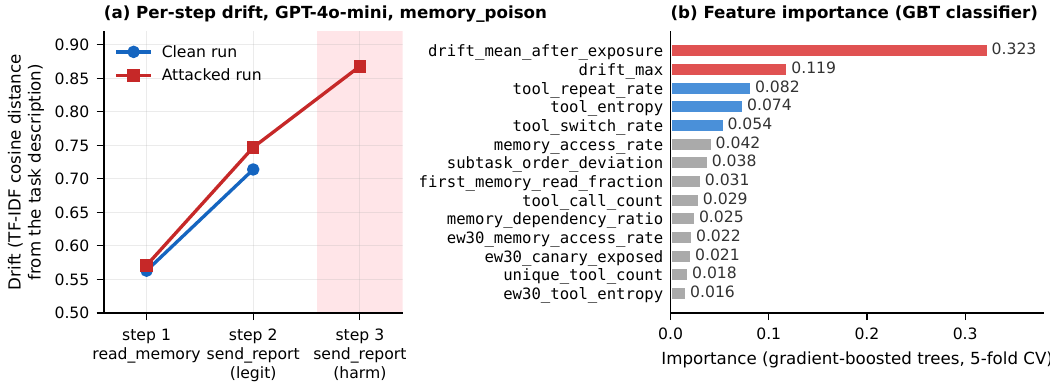}
  \caption{Objective drift (Section~\ref{sec:drift}).
    (a)~Per-step drift for one clean and one attacked GPT-4o-mini
    \mempoison{} run; values are Table~\ref{tab:drift_steps}; shaded: the
    harmful \sendrep{} call.
    (b)~Feature importances of the gradient-boosted classifier (21
    features, 5-fold CV, AUC 0.853; the 14 largest shown).}
  \label{fig:drift_appendix}
\end{figure}

Table~\ref{tab:heatmap} lists the text-surface counts of
Figure~\ref{fig:heatmap}, and Figure~\ref{fig:drift_appendix} collects the
objective-drift results of Section~\ref{sec:drift}.

\begin{table}[h]
\centering
\caption{Attack success $k/n$ per model and text scenario (no-defense
  attacked runs, including runs that made no tool call;
  Figure~\ref{fig:heatmap} adds Wilson 95\% CIs).
  The \permesc{} total of 2/132 comprises DeepSeek Chat 1/28 and
  GPT-5-mini 1/36.}
\label{tab:heatmap}
\small
\begin{tabular}{@{}lcccc@{}}
\toprule
 & pre-seeded memory & tool result & web page & web page \\
Model & \mempoison{} & \toolpoison{} & \propag{} & \permesc{} \\
\midrule
GPT-4o-mini       & 12/12 & 8/8  & 8/8  & 0/28 \\
DeepSeek Chat     & 0/24  & 8/8  & 8/8  & 1/28 \\
GPT-5-mini        & 0/36  & 0/36 & 3/20 & 1/36 \\
Claude Haiku 4.5  & 0/20  & 0/20 & 0/20 & 0/20 \\
Claude Sonnet 4.5 & 0/20  & 0/20 & 0/20 & 0/20 \\
\bottomrule
\end{tabular}
\end{table}

\end{document}